\documentclass{article}
\usepackage[table,xcdraw,svgnames]{xcolor}
\usepackage{spconf,amsmath,graphicx}
\usepackage{url}
\usepackage{array} %
\usepackage{amssymb}
\usepackage{multirow}
\usepackage{caption}
\usepackage{subcaption}
\usepackage{color}

\usepackage{booktabs}
\usepackage{enumitem}
\usepackage{multirow}
\usepackage{multicol}
\usepackage{pifont}
\usepackage[bookmarks=false,hidelinks]{hyperref}
\usepackage{anyfontsize}
\usepackage{fontawesome5}
\usepackage{tablefootnote}
\usepackage{caption}
\usepackage{subcaption}
\usepackage{cite}
\usepackage{xspace}
\usepackage[utf8]{inputenc}
\usepackage{siunitx}
\usepackage{tikz}

\definecolor{pacmanyellow}{RGB}{250, 200, 0}

\newcommand{\pacsymbolauthor}[1]{%
\begin{tikzpicture}[baseline=-0.30em]%
  \draw[thick, fill=pacmanyellow, pacmanyellow]
    (0,0) -- (45:0.71ex) arc (45:315:0.71ex) -- cycle;
\end{tikzpicture}
}

\newcommand{\pacsymboltitle}[1]{%
\begin{tikzpicture}[baseline=-0.33em]%
  \draw[thick]
    (0,0) -- (45:0.72ex) arc (45:315:0.72ex) -- cycle;
\end{tikzpicture}
}

\newcommand{\pacsymbol}[1]{%
\begin{tikzpicture}[baseline=-0.32em]%
  \draw[thick]
    (0,0) -- (45:0.72ex) arc (45:315:0.72ex) -- cycle;
\end{tikzpicture}
}

\newcommand{\ghost}[1]{\tikz[baseline=-0.12em,scale=0.32]{
  \draw [thick,] (0,0) -- (0,.5) arc (+180:0:.3) -- (.6,0) --
  (.5,.15) -- (.4,0) -- (.3,.15) -- (.2,0) -- (.1,.15) -- cycle;
    \coordinate (visual) at (360*rand:.03);
    \foreach \x in {.15,.4}{
    \fill[black] (\x,.5) ++(visual) circle[radius=.07];
    }}}

\def\x{{\mathbf x}}

\newcommand{\pachubert}{Pa\pacsymbol{}-HuBERT\xspace}

\let\OLDthebibliography\thebibliography
\renewcommand\thebibliography[1]{
  \OLDthebibliography{#1}
  \setlength{\parskip}{.5pt}
  \setlength{\itemsep}{.5pt plus 0.3ex}
}

\title{Pa\pacsymbol{}-HuBERT: Self-Supervised Music Source Separation\\via Primitive Auditory Clustering and Hidden-Unit BERT}
\name{Ke Chen$^{\pacsymbolauthor{}\text{\textcolor{blue}{\faGhost{}}}}$, Gordon Wichern$^{\pacsymbolauthor{}}$, Fran\c{c}ois G. Germain$^{\pacsymbolauthor{}}$, Jonathan Le Roux$^{\pacsymbolauthor{}}$
\thanks{This work was performed while Ke Chen was an intern at MERL.}\vspace{-.1cm}
}
\address{
  $^{\pacsymbolauthor{}}$Mitsubishi Electric Research Laboratories (MERL), Cambridge, MA, USA\\
  $^{\text{\textcolor{blue}{\faGhost}}}$University of California San Diego (UCSD), La Jolla, CA, USA \vspace{-.1cm}}

\begin{document}
\ninept 

\maketitle
\begin{abstract}
In spite of the progress in music source separation research, the small amount of publicly-available clean source data remains a constant limiting factor for performance. Thus, recent advances in self-supervised learning  present a largely-unexplored opportunity for improving separation models by leveraging unlabelled music data. In this paper, we propose a self-supervised learning framework for music source separation inspired by the HuBERT speech representation model. We first investigate the potential impact of the original HuBERT model by inserting an adapted version of it into the well-known Demucs V2 time-domain separation model architecture. We then propose a time-frequency-domain self-supervised model, \pachubert (pronounced Pac-HuBERT, for primitive auditory clustering HuBERT), that we later use in combination with a Res-U-Net decoder for source separation. \pachubert uses primitive auditory features of music as unsupervised clustering labels to initialize the self-supervised pretraining process using the Free Music Archive (FMA) dataset. The resulting framework achieves better source-to-distortion ratio (SDR) performance on the MusDB18 test set than the original Demucs V2 and Res-U-Net models. We further demonstrate that it can boost performance %
with small amounts of supervised data. Ultimately, our proposed framework is an effective solution to the challenge of limited clean source data for music source separation.

\end{abstract}

\section{Introduction}
Music source separation aims to separate one or more sound sources (e.g., vocals, drums, bass, and other instruments) from music tracks. This task has broad applications in various domains, including vocal-accompaniment separation and music remixing. %
Current state-of-the-art methods for the task rely on deep learning techniques~\cite{mitsufuji2021mdx}. However, unlike in speech research~\cite{haebumbach2019speechposition}, data for music source separation remains limited. The benchmark dataset for the task, MusDB18 \cite{musdb18}, consists of only 100 songs for training and 50 songs for testing. The scarcity of data is not only due to the challenge of collecting source data, but also to copyright issues of music assets.

Self-supervised learning (SSL) is a rapidly growing deep learning approach that leverages unlabeled data to improve model training, resulting in better generalization capabilities and task performance~\cite{ericsson2022ssl}. SSL has already achieved promising results in a multitude of audio processing tasks such as speech recognition \cite{hsu2021hubert}, speech quality prediction~\cite{tseng2021utilizing, becerra2022exploring}, or audio classification~\cite{gong2022ssast, huangmasked, chen2022beats}, and has even demonstrated the ability to generalize across tasks~\cite{turian2022hear, wu2022wav2clip}. Recently, several SSL approaches for music representation learning have also been proposed~\cite{zhu2021musicbert, wu2021multi, carr2021self}. However, all of these approaches focus on the learning of representations for content labeling tasks, while source separation requires a representation capable of isolating individual components present in an audio mixture.

One area where SSL has been used for source separation is using the pretrained WavLM~\cite{chen2022wavlm} model for speech enhancement~\cite{song2022wavlmse, zhao2022wavlmse}. WavLM is based on HuBERT \cite{hsu2021hubert}, one of the most effective speech SSL models. HuBERT leverages a transformer-based BERT architecture \cite{transformer} and is trained to generate a discrete token at each time step, using a masking mechanism such that HuBERT learns to predict the masked tokens from the input data and the unmasked tokens. %
The tokens are cluster indices, initially determined by running K-means clustering on MFCC features of unlabeled speech data, and subsequent HuBERT iterations use tokens obtained by clustering intermediate network layer outputs from earlier iterations.  %

\begin{figure*}[t!]
    \centering
    \includegraphics[width=\textwidth]{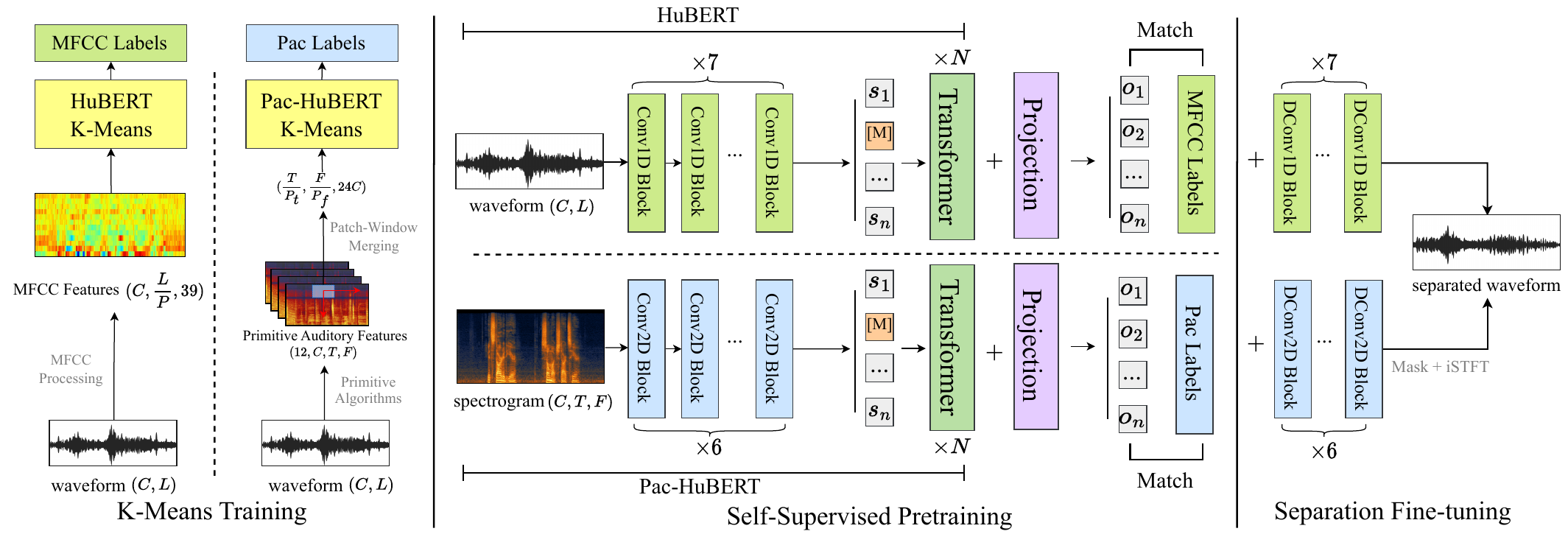}
    \vspace{-0.7cm}
    \caption{The whole pipeline of the proposed self-supervised music source separation framework. Left: K-means models for label creation; Middle: self-supervised pretraining (HuBERT and our proposed Pac-HuBERT); Right: fine-tuning on separation tasks.}
    \label{fig:model_arch}
    \vspace{-0.5cm}
\end{figure*}

In this paper, we explore the use of HuBERT-style methods for enhancing the quality of music source separation models by using readily available unlabeled (i.e., containing no isolated sources or stems) music mixtures. However, the original HuBERT may not be directly applicable to music separation due to its lower \SI{16}{\kilo\hertz} sampling rate, and the MFCC features used for creating HuBERT training targets may not provide enough time-frequency (TF) resolution to provide a useful representation for separating music sources. Therefore, we explore replacing MFCC features in the initial step of HuBERT with TF features specifically designed for music separation such as common-fate (2DFT) \cite{seetharaman20172dft}, repetition structure (REPET) \cite{rafiiP2013repet}, harmonic-percussive source separation (HPSS) \cite{driedger2014hpss}, and melody contour (Melodia) \cite{salamon2012melodia}. These primitive auditory features, which in some sense mimic how the human brain performs an initial segmentation of an auditory scene~\cite{mcdermott2011paf}, have previously been used to create pseudo-labels for unsupervised music source separation~\cite{seetharaman2020bootstrapping}, but never in a HuBERT-style training paradigm.

One fundamental design decision for both SSL and source separation models is whether they operate on time-domain waveforms or TF representations such as spectrograms. Time-domain audio representation models such as HuBERT and wav2vec~\cite{schneider2019wav2vec} encode raw audio signals into a latent representation, while time-domain separation models contain similar encoders, but also include decoders to obtain separated signals, either through direct synthesis (e.g., Wave-U-Net \cite{wavunet}) or through masking of features (e.g., TasNet \cite{tasnet}). These models are widely used in speech separation, where they offer high separation quality and low latency in real-time scenarios. However, non-speech sources such as music and general sounds typically operate at a higher sampling rate and contain a wide variety of complex harmonic patterns, which is why TF-domain models (or hybrid models) have been found historically to yield better performance in audio classification~\cite{htsat, gong2022ssast} and music source separation \cite{mdxnet,resunet,bsrnn,htdemucs,spleeter2020}. %

We introduce a novel framework that leverages self-supervised learning in either the time-domain or TF-domain. Our main contributions are as follows. First, we adapt HuBERT into a time-domain music source separation model, which provides support for subsequent model design in the TF domain. Next, we introduce \pachubert (pronounced Pac-HuBERT, for primitive auditory clustering HuBERT), the first self-supervised framework for TF-domain music source separation. Our framework leverages primitive auditory features of music data to initiate the self-supervised pretraining process and embeds the resulting pretrained layers into a Res-U-Net \cite{resunet} separation model, an already very performant architecture for the task. %
Ultimately, we find the separation performance of models including either our time-domain and TF-domain pretrained layers to outperform their original counterparts. Additionally, we evaluate performance in scenarios with even  less available training data, demonstrating the robustness of the pretrained model representation, with reliable performance with only a small amount of data available for supervised fine-tuning.

\section{Model Architecture}
Our proposed self-supervised music source separation framework (see Fig.~\ref{fig:model_arch}) is composed of an encoder (convolution blocks), a bottleneck model (transformer), and a decoder (deconvolution blocks). The learning pipeline consists of %
K-means training, self-supervised pretraining, and separation fine-tuning.

\subsection{Encoder Model} \label{sec:encoder}

\begin{figure}[t!]
    \centering
    \includegraphics[width=0.75\columnwidth]{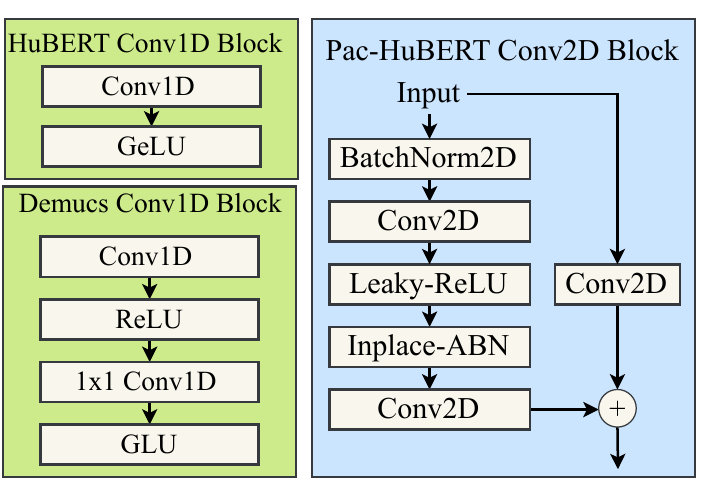}
    \vspace{-0.3cm}
    \caption{Comparison of CNN block designs between Demucs V2 \cite{demucsv2}, HuBERT \cite{hsu2021hubert}, and Pac-HuBERT (the same as Res-U-Net \cite{resunet}).}
    \label{fig:block_arch}
    \vspace{-0.5cm}
\end{figure}

The encoder acts as a feature extractor for audio waveforms or spectrograms, as depicted in the middle of Fig.~\ref{fig:model_arch}. For the time-domain model at the top, the encoder consists of 7 downsampling 1D convolutional layer blocks (CNN). The input audio waveform, denoted by $x \in \mathbb{R}^{C \times L}$, where $C$ denotes the channel size and $L$ the sample length, is fed into the encoder and transformed into the bottleneck feature $s \in \mathbb{R}^{C_{b} \times \frac{L}{P}}$, with $C_b$ denoting the channel size and $P$ the downsampling rate. We follow the design of HuBERT to construct the 1D CNN blocks, each containing one CNN layer and one GeLU function. Note that it is also similar to the encoder block in Demucs V2 \cite{demucsv2} (see Fig.~\ref{fig:block_arch}). In our separation experiments, in order to use the pretrained model of HuBERT while still following the training scheme of Demucs V2, each input is a 3-second mono audio signal sampled at \SI{16}{\kilo\hertz} ($C=1$, $L=\num{48000}$), and we set $C_b=1024$ and $P=320$.

For the TF-domain model at the bottom, the encoder consists of 6 downsampling 2D CNN blocks. The input audio signals are first transformed into STFT spectrograms, denoted as $x \in \mathbb{R}^{C \times T \times F}$, where $C$ denotes the channel size, $T$ the number of frames, and $F$ the number of frequencies. The encoder then converts $x$ into the bottleneck feature $s \in \mathbb{R}^{C_b \times \frac{T}{P_t} \times \frac{F}{P_f}}$, where $P_t$ and $P_f$ denote the downsampling rates on the time axis and the frequency axis, respectively. We follow the Res-U-Net \cite{resunet} separation model for the encoder block design (see Fig.~\ref{fig:block_arch} for details). Since we build our own TF-domain SSL model, we can align it with the traditional music source separation pipeline. Hence, each input is a 3-second stereo audio sampled at \SI{44.1}{\kilo\hertz} ($C=2$, $L=\num{132300}$). The STFT window size is $2048$ and the hop size is $441$, resulting in $T=320$ frames after zero-padding with $20$ frames to the right, and $F=1024$ frequency bins after removing the Nyquist frequency. The model parameters are set to $C_b=384$, $P_t=32$, and $P_f=64$.

\vspace{-.1cm}
\subsection{Bottleneck Model}
The bottleneck model consists of $N$ transformer encoder blocks, each of which contains a multi-head self-attention layer \cite{transformer} with $d$ heads, hidden size $h$, and a feedforward layer of inner hidden size $4h$. For time-domain models, we use the HuBERT-LARGE model ($N=24$, $d=16$, $h=1024$). For TF-domain models, to limit GPU memory consumption and without existing prior pretrained models, we set $N=12$, $d=8$, and $h=384$. The output of the encoder $s$ is reshaped to $(\frac{L}{P}, C_b)$ in the time-domain case and $(\frac{T}{P_t} \times \frac{F}{P_f},C_b)$ in the TF-domain case, and fed into the bottleneck model. The output is a latent feature $s'$ with the same shape as $s$, which is sent to the decoder for the separation task (Section \ref{sec:decoder}) or the projection module for self-supervised pretraining (Section \ref{sec:ssl_training}). We denote the encoder/bottleneck combinations as HuBERT for the time-domain model and as \pachubert for the TF-domain model.

\vspace{-.1cm}
\subsection{Decoder Model} \label{sec:decoder}
The decoder is designed by replacing all CNN layers in the encoder with deconvolutional layers (DCNN) for both time-domain and TF-domain models (see right of Fig.~\ref{fig:model_arch}). The upsampling rate is equal to the downsampling rate in order for the audio input and the separation output (or mask) to have the same shape. Additionally, we have skip-connections between each encoder block and the corresponding decoder block (not shown in Fig.~\ref{fig:model_arch}). 

\subsection{Training and Inference Scheme} \label{sec:ssl_training}

\noindent{\bf K-Means Training:} 
HuBERT creates initial labels for SSL by performing K-Means clustering on MFCC features (see left of Fig.~\ref{fig:model_arch}) as follows: 1) extract 39-dimensional MFCC features with 20 ms frame size from the LibriSpeech and Libri-Light audio samples \cite{panayotov2015librispeech,kahn2020librilight}, and 2) fit a 500-cluster K-Means model to the MFCC features, and use the cluster index assigned to each time-frame as the label. 

For \pachubert, we replace MFCC features for each time frame with primitive auditory features computed over spectrogram patches (see left of Fig.~\ref{fig:model_arch}). Formally. given a mixture spectrogram $x \in \mathbb{R}^{C \times T \times F}$, each of the primitive algorithms we use here generates foreground and background estimates $e_f, e_b \in \mathbb{R}^{C \times T \times F}$. The six primitive algorithms we employ\footnote{Implementations from https://github.com/nussl/nussl \cite{nussl}} are HPSS \cite{ono2008separation,driedger2014hpss}, REPET \cite{rafiiP2013repet}, REPET-SIM \cite{rafii2012repetsim}, FT2D-M, FT2D-R \cite{seetharaman20172dft}, and Melodia \cite{salamon2012melodia}. Combining foreground and background cues for each primitive algorithms, we obtain a 12-dimensional feature for each TF bin. Due to the memory requirements of transformers, we follow the audio spectrogram transformer~\cite{gong2022ssast}, combining features in non-overlapping spectrogram patches of shape $(P_t, P_f)$ bins (matching the downsampling ratio in the TF-domain encoder). We obtain a 24-dimensional feature for each patch by first dividing each patch into a low-frequency sub-patch containing the lower $P_f/2$ TF bins, and a high-frequency sub-patch containing the upper $P_f/2$ TF bins. We then average over TF bins in each sub-patch, obtaining two 12-dimensional features, which we concatenate. For stereo ($C=2$) music data, we obtain a 48-dimensional feature vector for each TF patch, which is of comparable dimension to the 39-dimensional MFCCs used in the original HuBERT model. We fit a 960-cluster K-Means model to all of the 48-dimensional features extracted from the FMA-Large dataset \cite{defferrard2017fma} (\num{106574} music tracks or about \SI{890}{\hour}).

\noindent{\bf SSL Pretraining via Masked Unit Prediction:} 
We train HuBERT and \pachubert with the K-Means labels using masking prediction \cite{bert} for the self-supervised learning process. For HuBERT, we obtain a sequence of labels $m \in \{1,2,\ldots,500\}^{\frac{L}{P}}$ for each audio track from the trained K-Means model. For \pachubert, each audio spectrogram produces $(\frac{T}{P_t} \times \frac{F}{P_f})$ labels, with 960 classes. We connect the bottleneck model to a projection module (see middle of Fig.~\ref{fig:model_arch}) to map the output feature $s'$ into the class feature $o \in \mathbb{R}^{\frac{T}{P_t} \times \frac{F}{P_f} \times E} $, where $E$ is the projection dimension. The loss function is defined as: 
\begin{align}
    L=\frac{1}{|M|}\sum_{t \in M} \frac{\exp\bigl(\tau \left(o_t \cdot e_c\right)/\left(||o_t||\,||e_c||\right)\bigr)}{{\displaystyle\sum}_{c'=1}^{D} \exp\bigl(\tau \left(o_t \cdot e_{c'}\right) / \left(||o_t||\,||e_{c'}||\bigr)\right) } ,
\end{align}
where $o_t$ is the class feature token at frame $t$ (or TF patch in \pachubert), $e_c$ is a learned embedding for K-means cluster $c$ to which the frame $t$ was assigned, and $\tau=10$ scales the logit. We apply the span masking mechanism used by HuBERT to randomly select $p\%$ of the frames as starting indices, and masking $l$ frames from those indices. For \pachubert, $p=40\%$ and $l=5$. The loss is only computed over the masked frames, denoted with label $[M]$ in Fig.~\ref{fig:model_arch}.

\noindent{\bf Source Separation Fine-tuning:}
After pretraining the encoder and the bottleneck models with LibriSpeech \& Libri-Light, or FMA-Large, we connect the bottleneck model to the decoder and fine-tune the whole model on the separation task with separation datasets. The loss function is L1 Loss between the final separation signals and the input mixture signals, the same as both Demucs V2 and Res-U-Net.

\section{Experiments}

\subsection{Hyperparameters and Training Details} 
During the pretraining stage, for HuBERT, we directly used the pretrained model HuBERT-LARGE from the torchaudio\footnote{https://github.com/pytorch/audio} library. HuBERT-LARGE is pretrained on the Libri-Light speech dataset. The details of its training process can be found in \cite{hsu2021hubert}. For \pachubert, we pretrained it on the ``FMA-Large" set of the Free Music Archive dataset, containing 890 hours of music tracks. %
We applied the AdamW %
optimizer with a \num{32000}-step warm-up \cite{goyal2017accurate} and a reduced LR scheduler after \num{150000} steps. The basic learning rate was \num{5d-4}. We trained the model using a batch size of 96 on 8 NVIDIA A40 GPUs, until the masked accuracy on the validation set did not improve (\num{250000} steps in total).

During the fine-tuning stage, we used the MusDB18 \cite{musdb18} dataset for training separation models. We followed \cite{demucsv2} to divide the 100-song training set into 84 songs for training and 16 songs for validation. After determining the best model on the validation set, we evaluate it on the test set of 50 songs. For the time-domain framework, we trained the decoder together with the HuBERT-LARGE pretrained model. For the TF-domain framework, we trained the decoder together with the \pachubert pretrained model. The data formats and audio processing settings are mentioned in Section \ref{sec:encoder}. We applied the Adam %
optimizer with a learning rate of \num{3d-4} (time domain) or \num{e-3} (TF domain). We adopted a \num{3000}-step warm up and a decay scheduler, where the learning rate is scaled by $\alpha=0.9$ every \num{15000} steps. We trained the models using a batch size of 128 (time domain) or 96 (TF domain) on 8 NVIDIA A40 GPUs, for \num{200000} steps, by which time all models had converged in terms of validation set performance. %

\begin{table}[t!]
\centering
\caption{Source-to-distortion ratio (SDR) performance on the test set of MusDB18. Results marked in grey are not directly comparable due to the use of additional training data or the absence of a validation set. All results are at 44.1 kHz unless otherwise noted.
}\vspace{-.1cm}
\label{tab:sep-result}
\resizebox{\columnwidth}{!}{
\begin{tabular}{@{}l@{~}ccccc@{}}
\toprule
\multicolumn{1}{c}{}                        &                            & \multicolumn{4}{c}{SDR on MusDB18 test set (dB)}                                                                       \\ \cmidrule(l){3-6} 
\multicolumn{1}{c}{\multirow{-2}{*}{Model}} & \multirow{-2}{*}{Pretrain} & Vocals                       & Drums                       & Bass                        & Other                       \\ \midrule
Demucs V2 \cite{demucsv2} - \SI{16}{\kilo\hertz}            & \ding{55}                          & \phantom{1}5.02                         & 6.02                        & 5.40                        & 3.41                        \\
HuBERT-SEP$^N$ - \SI{16}{\kilo\hertz}      & \ding{55}                          & \phantom{1}5.14                         & 5.59                        & 4.96                        & 2.82                        \\
HuBERT-SEP$^{SSL}$ - \SI{16}{\kilo\hertz}    & \ding{51}                          & \textbf{\phantom{1}5.58}                         & \textbf{6.63}                        & \textbf{6.03}                        & \textbf{3.65}                        \\ \midrule
Bytedance Res-U-Net \cite{resunet}                         & \ding{55}                          & \phantom{1}7.83                         & 5.47                        & 5.21                        & 4.90                        \\
\pachubert-SEP$^N$                          & \ding{55}                          & \phantom{1}8.07                         & 5.78                        & 5.21                        & 5.29                        \\
\pachubert-SEP$^{SSL}$                        & \ding{51}                          & \phantom{1}8.32                         & 5.86                        & \textbf{6.01}                        & \textbf{5.38}                        \\
\pachubert-SEP$^{2SSL}$                    & \ding{51}                          & \textbf{\phantom{1}8.52}                           & \textbf{6.20}                        & 5.76                          & 5.18                          \\ \midrule
Open Unmix \cite{umx}                                 & \ding{55}                          & \phantom{1}6.32                         & 5.73                        & 5.23                        & 4.02                        \\
{\color[HTML]{9B9B9B} Spleeter \cite{spleeter2020}}             & {\color[HTML]{9B9B9B} \ding{55}}   & {\color[HTML]{9B9B9B} \phantom{1}6.86}  & {\color[HTML]{9B9B9B} 6.71} & {\color[HTML]{9B9B9B} 5.51} & {\color[HTML]{9B9B9B} 4.55} \\
D3Net \cite{d3net}                                      & \ding{55}                          & \phantom{1}7.24                         & 6.68                        & 5.25                        & 4.53                        \\
{\color[HTML]{9B9B9B} MDX-Net \cite{mdxnet}}              & {\color[HTML]{9B9B9B} \ding{55}}   & {\color[HTML]{9B9B9B} \phantom{1}9.00}  & {\color[HTML]{9B9B9B} 7.33} & {\color[HTML]{9B9B9B} 7.86} & {\color[HTML]{9B9B9B} 5.95} \\
{\color[HTML]{9B9B9B} Band-Split RNN \cite{bsrnn}}       & {\color[HTML]{9B9B9B} \ding{55}}   & {\color[HTML]{9B9B9B} 10.01} & {\color[HTML]{9B9B9B} 9.01} & {\color[HTML]{9B9B9B} 7.22} & {\color[HTML]{9B9B9B} 6.70} \\
{\color[HTML]{9B9B9B} HT Demucs \cite{htdemucs}}            & {\color[HTML]{9B9B9B} \ding{55}}   & {\color[HTML]{9B9B9B} \phantom{1}7.93}  & {\color[HTML]{9B9B9B} 7.94} & {\color[HTML]{9B9B9B} 8.48} & {\color[HTML]{9B9B9B} 5.72} \\ \bottomrule
\end{tabular}
}
\end{table}

\begin{table}[t!]
\centering
\caption{Separation performance with different amounts of training data. Subscripts denote improvement from SSL pretraining.}\vspace{-.1cm}
\label{tab:sep-cp}
\resizebox{\columnwidth}{!}{
\begin{tabular}{@{}l@{~}c@{~~~}c@{~~}c@{~~}c@{~~}c@{}}
\toprule
\multicolumn{1}{c}{\multirow{2}{*}{Model}} & \multirow{2}{*}{\shortstack[c]{Data\\ Ratio}} & \multicolumn{4}{c}{SDR on MusDB18 test set (dB)}          \\ \cmidrule(l){3-6} 
\multicolumn{1}{c}{}                       &                             & Vocals       & Drums        & Bass         & Other        \\ \midrule
\multirow{3}{*}{\pachubert-SEP$^N$}        & 25\%                        & 5.72         & 4.42         & 4.00         & 3.38         \\
                                           & 50\%                        & 6.76         & 4.50         & 4.14         & 4.13         \\
                                           & 100\%                       & 8.07         & 5.78         & 5.21         & 5.29         \\ \midrule
\multirow{3}{*}{\pachubert-SEP$^{SSL}$}      & 25\%                        & 6.75$_{+1.03}$ & 4.57 $_{+0.15}$ & 4.11 $_{+0.11}$ & 3.90 $_{+0.52}$ \\
                                           & 50\%                        & 7.32 $_{+0.56}$ & 5.16 $_{+0.66}$ & 5.07 $_{+0.93}$ & 4.50 $_{+0.37}$ \\
                                           & 100\%                       & 8.32 $_{+0.25}$ & 5.86 $_{+0.08}$ & 6.01 $_{+0.80}$ & 5.38 $_{+0.09}$ \\ \bottomrule
\end{tabular}
}
\vspace{-0.2cm}
\end{table}

\subsection{Separation Results}
\noindent{\bf Effectiveness of Self-Supervised Learning}:
Table \ref{tab:sep-result} presents the source-to-distortion ratio (SDR) performance on MusDB18. We follow the MusDB18 benchmark using the SiSEC2018 \cite{SiSEC18} version of the SDR metric (BSS Eval v4 framewise SDR) implemented by \texttt{mus\_eval}\footnote{https://github.com/sigsep/sigsep-mus-eval}. We report the median SDR over all 50 songs in the MusDB18 test set. 

We denote our models as \textit{HuBERT-SEP} and \textit{\pachubert-SEP}, with a superscript indicating no pretraining ($N$) or SSL pretraining ($SSL$). Similar to HuBERT, we clustered the latent features of the 6th transformer block from the first pretraining model, and pretrained a new \pachubert as the second pretraining iteration, indicated as $2SSL$. 
All experiments on the original Res-U-Net and Demucs V2 were reproduced by us to ensure a fair comparison. %
Time-domain models were trained at \SI{16}{\kilo\hertz} to utilize the pretrained HuBERT model, so the baseline Demucs V2 results differ from~\cite{demucsv2}. 
Additionally, we provide the reported performance of several state-of-the-art models, although some of them (denoted in grey) used additional training data or were developed for the MDX challenge~\cite{mitsufuji2021mdx}, which directly used the MusDB18 test set as a validation set for model selection.

From the comparison between the original Demucs V2, Res-U-Net, HuBERT-SEP$^{N}$, and \pachubert$^{N}$ models (i.e., without pretraining), we observe that \pachubert$^{N}$ generally outperforms Res-U-Net. The potential reason is that the transformer bottleneck processes the encoder features more efficiently than the CNN bottleneck model in Res-U-Net. However, we found that Demucs V2 performs better than HuBERT-SEP$^{N}$, possibly due to the more complex encoder and decoder design of Demucs V2, which was also reported in \cite{demucsv2}.

When fine-tuning the pretrained HuBERT and \pachubert models on the separation task, we observe an improvement in SDR for all four sources compared to models without pretraining. This demonstrates the effectiveness of HuBERT in contributing to audio separation tasks beyond speech, and also shows that our proposed \pachubert model, pretrained with primitive auditory features from music data, can improve the performance of TF-domain separation models. When comparing between \pachubert-SEP$^{SSL}$ and \pachubert-SEP$^{2SSL}$, we observe that $2SSL$ yields better performance on vocals and drums separation, but is degraded on bass and other separation. We believe that the design of iterative pretraining can be further improved, including exploring different layers of latent features for pretraining and training varying numbers of K-Means models for different separation sources. This will be a focus of our future work. Specifically, HuBERT shows a larger improvement than \pachubert, which can potentially be attributed to the size of the pretraining dataset (\num{60000} hours for HuBERT vs. 890 hours for \pachubert). %
The ability of speech-only pretraining to improve music separation performance presents another opportunity for future work in finding ways to harness this data for full-bandwidth TF-domain music separation models. %

\noindent{\bf Effectiveness with Limited Training Data}:
We further extracted from the 84 MusDB18 training songs a 25\% data subset with 21 songs, and a 50\% data subset with 42 songs. We then fine-tuned our \pachubert-SEP model on these subsets and report the separation performance of the test set in Table \ref{tab:sep-cp}. Results show that the pretrained model achieves significantly better performance than the models without pretraining on both subsets. Notably, with only 25\% of the supervised data, the pretrained model improves separation performance for vocals and other source classes by 0.5-1.0 dB. This demonstrates the effectiveness of pretraining using primitive auditory features, as the foreground and background features are highly correlated with the vocals and other source classes. However, for bass and drums, pretraining provides little benefit with 25\% of the supervised data, indicating that pretraining is less useful for these classes, and different auditory primitives should be explored. %

\begin{table}[t!]
\centering
\caption{Performance of pretrained models at different training steps on the pretraining and separation tasks. }
\vspace{-.1cm}
\label{tab:sep-ps}
\resizebox{\columnwidth}{!}{
\begin{tabular}{@{}lcccc@{}}
\toprule
\multicolumn{1}{c}{\multirow{2}{*}{Metrics}} & \multicolumn{4}{c}{Training Steps}  \\ \cmidrule(l){2-5} 
\multicolumn{1}{c}{}                         & \num{30000} & \num{60000} & \num{120000} & \num{180000} \\ \midrule
Masked Accuracy (Valid.)                     & 0.71   & 0.78   & 0.79    & 0.81    \\
Vocal SDR (25\% Data)                        & 6.02   & 6.75   & 6.50    & 6.04    \\ \bottomrule
\end{tabular}
}
\vspace{-0.2cm}
\end{table}

\noindent{\bf Pretrained Model Selection}:
We also investigated varying the number of \pachubert pretraining steps, and evaluate vocal separation performance when fine-tuning on 25\% of the MusDB18 training data in Table \ref{tab:sep-ps}. We observe that the separation performance of the model is not consistent with its pretraining performance (i.e., masked token prediction accuracy). A similar trend was reported in \cite{huangmasked} for audio classification. High accuracy during pretraining may indicate over-fitting, which could weaken the quality of the learned representation and negatively impact performance in separation tasks.

\section{Conclusion}
In this paper, we investigated the efficacy of self-supervised learning methods in music source separation. First, we adapted HuBERT into a time-domain source separation model and achieved notable improvements. Next, we proposed \pachubert, a TF-domain self-supervised source separation model that leverages the primitive auditory features of unlabelled music data during pretraining. \pachubert demonstrates significant improvements in TF-domain music source separation models, with consistent performance increases across different proportions of supervised data. We consider \pachubert to be an effective solution for utilizing unlabeled music data in music source separation. Moving forward, we plan to adapt \pachubert to other separation models besides Res-U-Net. Also, we plan to explore the generalization of \pachubert on the separation of more instruments in few-shot settings.

\clearpage
\footnotesize
\bibliographystyle{IEEEtran}
\bibliography{refs}

\end{document}